\documentclass[10pt, conference, compsocconf]{IEEEtran}
\ifCLASSINFOpdf
   \usepackage[pdftex]{graphicx}
   \DeclareGraphicsExtensions{.pdf,.jpeg,.png}
\else
   \usepackage[dvips]{graphicx}
\fi
\usepackage[cmex10]{amsmath}
\begin{document}
%
\title{
Can an Ad-hoc ontology Beat a Medical Search Engine? The Chronious Search Engine case.
}


\author{%
Piero Giacomelli \\
Tesan S.p.A., Italy \\
email:giacomelli@tesan.it \\
 \and
Giulia Munaro\footnote{munaro@tesan.it} \\
Tesan S.p.A., Italy \\
email:munaro@tesan.it \\
 \and
Roberto Rosso\footnote{rosso@tesan.it}   \\
Tesan S.p.A., Italy \\
email:rosso@tesan.it \\
           }

%


\maketitle

\begin{abstract}
		Chronious is an Open, Ubiquitous and Adaptive Chronic Disease Management Platform for Chronic Obstructive Pulmonary Disease(COPD)
	Chronic Kidney Disease (CKD) and Renal Insufficiency. It consists of several modules: an ontology based literature search engine, a rule based decision support system, remote sensors interacting with lifestyle interfaces (PDA, monitor touch-screen) and a machine learning module. All these modules interact each other to allow the monitoring of two types of chronic diseases and to help clinician in taking decision for care purpose.
	This  paper illustrates how the ontology search engine was created and fed and how some comparative test indicated that the ontology based approach give better results, on some estimation parameters, than the main reference web search engine.
\end{abstract}

\begin{IEEEkeywords}
\begin{bfseries}
\begin{itshape}
Telemedicine; chronic disease management; ontology search engine. 
\end{itshape}
\end{bfseries}
\end{IEEEkeywords}

%
\IEEEpeerreviewmaketitle

\section{Introduction}

Scientific advances over the past 150 years, particularly in the medical field, have allowed the extension of life expectancy in western countries and this trend seems to increase in future years. Conservative estimates suggest that by 2030 in EU countries the proportion of people over 60 years  regard the entire population will be around 50$\%$; this means that we will see a gradual increase in the number of those subjects with chronic diseases (i.e., diseases not involving healing), that will therefore increase the cost and effort over health care facilities \cite{CarmineZoccali06012010}.

Chronic diseases are slowing but constantly replacing malnutrition and infection as primary causes of mortality in the population \cite{MILQ:MILQ398}. The World Health Organization (WHO) has recently emphasized that chronic diseases are a global priority \cite{WHO}.It was calculated that, if governments are able to put in place public health policies that produce a 2$\%$ yearly reduction in mortality rates for chronic diseases, 36 million deaths would be prevented worldwide between 2005 and 2015 \cite{citeulike:1514400}.

\indent
Chronic diseases are difficult to treat and, apart from deaths, have collateral social impact that are becoming an economic emergency both in western and developing countries. As the number of patient with chronic diseases is rising there will be an increasing cost for hospitalization structure both public and private. Considering some specific diseases like Chronic Kidney Disease (CKD), sometimes there is, during the medical treatment, a non-return point from where the hospitalization is continuous as for dialysed people. The traditional approach consisting in periodic check-ups and periodic lab exams seems a model that won't be sustainable as the population gets older and the total number of patients with chronic diseases rises. At present the physician deals with an increasing number of chronic patients that are lowering the  periodic check-ups and so reducing the ability to prevent, if not death, worsening in patient's quality of life.
\newline
\indent
In the latest years, we have seen a tremendous growth in IT infrastructure, both from the hardware and communication capacity. Nowadays a common mobile phone is much more powerful in terms of hardware and software capacity than the first calculating machine that allowed the man to land on the moon forty years ago. The continuous growth of the World Wide Web (WWW) and, linked to this, the continuous growth in bandwidth capacity for data transmission allows to have cheaper and more widely available bandwidth, for larger portions of the population. 
\indent
As consequence of the exponential growth of hardware and software infrastructure it is possible to rethink the whole approach to the treatment of complex chronic diseases by limiting the hospitalization only to situations of severe worsening of patient condition. This was the original idea behind the EU funded Chronious project \cite{chronious}: constructing a generic platform to monitor, in an unobtrusive way, patient with chronic disease in two goals \cite{Vitacca01022009}:

\begin{itemize}
\renewcommand{\labelitemi}{$\bullet$}
\item{Improve the patients quality of life, by reducing as much as possible the hospitalizations.}
\item{Allow the clinician a continuous monitoring of the patients, both in standard and  potential risk situations.}
\end{itemize}

To gain this two goals, the Chronious platform has to integrate different technologies such as hardware and software modules that need to interact among themselves. This paper is focused on the ontology search engine module: we will illustrate what are the aims of this module and what are the main components.
The storage system for the documents is an ontology, developed specifically for the CODP and CKD diseases. We will illustrate how this ontology was created and enriched, from medical literature sources. Finally we will illustrate our tests conducted against the principal medical search engine and how the preliminary results seem to indicate that such approach can outperform the results of a web search engine. 
The paper is organized in the following sections. We will first describe the Ontology Chronious Search Module and what were the needs and how we solved them. Chapter three fully illustrate the whole process of document uploading and processing of the text to gain the enrichment of the ontology.
Finally we will illustrate our tests results and suggest some future improvements.

\section{Chronious Search Module}
Chronious is an hardware/software platform devoted to monitor in a remote way COPD and CKD patients. In 
Figure \ref{chronious}, a schema of the whole system is presented:

\begin{figure}[htbp]
	\centering
		\includegraphics[width=0.5\textwidth]{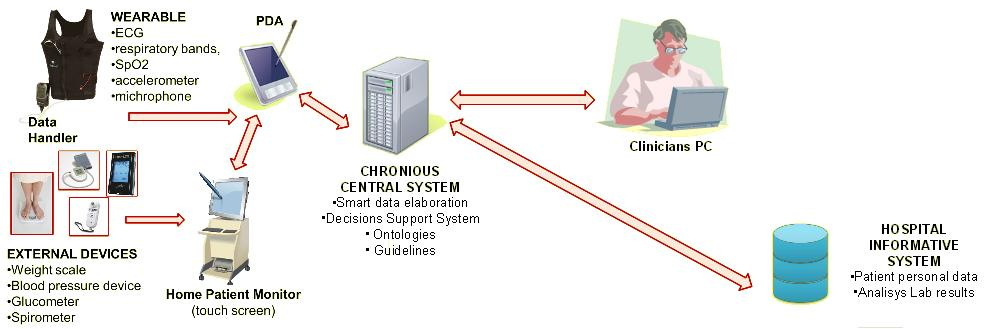}
	\caption{Chronious modules}
	\label{chronious}
\end{figure}

 The chronic patient is equipped at home with the following devices:
\begin{itemize}
\renewcommand{\labelitemi}{$\bullet$}
\item{a Personal Digital Assistant (PDA) that contains machine learning algorithms, acting as a first alerting system and that is used to transmit data to a central system through GPRS technology.}
\item{A Home Patient Monitor (HPM) that is used to allow patients to insert qualitative information on diet, activity and drugs intake.}
\item{Bluetooth medical devices: weight scale, blood pressure intake device, glucometer, air quality monitoring tool. Coupled with these devices a sensorized t-shirt for vital parameters recording like cardiac and respiratory signals.}
\end{itemize}
\indent
For the two pathologies, different sets of equipments are given to the patient according to the clinician. 
For example the glucometer is usually assigned to CKD patient affected by diabetes comorbidity .

The data collected are automatically transmitted via GPRS to a Central System that, using a web interface, and a rule-based Central Decision Support System (CDSS), allows clinician to monitor patient and to receive suggestions on how to act in case of a worsening trend or a potentially life risk situation. \newline
\indent
The CDSS uses JENA \cite{jena} framework and a set of rules codified in OWL \cite{OWL} format to display suggestions to the clinician. An example of such a rule is displayed in table \ref{tab:table2}.

\begin{table}[htbp]
	\label{tab:table2}
		\centering
	\caption{Example of a suggestion given by the CDSS}

\begin{tabular}{|l|l|}
\hline 
patientID & 1 \\ 
\hline 
AlertType & White \\ 
\hline 
Date & 01/02/2011 \\ 
\hline 
Description & body temperature up to 38$°$ \\ 
\hline 
Suggested action & hospitalization of the patient \\ 
\hline 
Guideline Text & text from literature	 \\
\hline 
\end{tabular} 

\end{table}
For CKD we used the KDOQI Guidelines \cite{citeulike:1541348}, for COPD we create a set of rules based on clinician experience.
The suggestions are portion of text extracted from literature reference or documents provided by the experts.
Managing documents and information , and organising concepts, such as comorbidities, and relations between them is a central task, for having a correct outcome to the CDSS calling. 
\indent
Searching literature reference, we find that latest information retrieval/storage systems studies, seem to indicate that ontology structure are a better way details concepts and the relations between them \cite{punitha}. Ontologies have been used successfully on medical \cite{Abasolo00melisa.an}, genetic \cite{citeulike:212874} and surgery \cite{conf/gi/MudunuriBN09} fields.  
The document repository for storing informations on the COPD and CKD diseases, was chooses as an ontology. This lead us to other issues: how to build the ontology and how to enrich it with new concepts and to validate it. 

\indent
Agreeing with doctors , the main reference on the medical field is the PubMed\cite{pubmed} site. PubMed is a free database accessing primarily the MEDLINE database of references and abstracts on life sciences and biomedical topics. As of 1 July 2011, PubMed has over $21$ million records going back to $1966$, selectively to the year $1865$, and very selectively to $1809$. About 500,000 new records are added each year. As of 1 July 2011, $11.9$ million articles are listed with their abstracts and $3.3$ million articles are available full-text for free.
\indent
 The problem was how to integrate all this huge among of information in the PubMed site with the Chronious Ontology. Apart from the integration the main problem was how to connect all sparse information about COPD and CKD pathologies in a way to produce a real value information to the clinician. Querying Pubmed for generic COPD (at $1$th September 2011), we have as outcome $33319$ documents. 

We coupled the Chronious Documents repository with a Search Module for having an interface that clinician can use to fast access literature specific information on the two diseases treated by Chronious.

Using these intuitions, the whole Chronious Search Module of the Chronious System is composed of four main parts:
\begin{itemize}
\renewcommand{\labelitemi}{$\bullet$}
\item{Upload tool: a web interface that is able to upload files to the repository.}
\item{The repository itself and the ontology used to underline concepts and relation extracted from the raw text.}
\item{Enrichment tool: a tool that after the text information extraction is able to say which new concepts and relations should be accepted.}
\item{A search tool. This is the final web interface for the clinician that using the concepts and terms provided in a free text search or more structured way is able to query the repository and give documents as feedback.}

\end{itemize}

Being that the Chronious Search Tool was the entry point for the CDSS,  the clinician feedbacks about the whole architecture underlined potential issues. The main issue, was that it is useless to use a personalized search engine while going via web to PubMed was so fast and so accessible. Another issue was that the clinician should upload manually the document into the repository. We will describe the processing of a single document to demonstrate how we solve these issues.

\section{The Chronious upload/processing system}
The Chronious ontology have been developed using the DOLCE ontology \cite{schneider}, however to enrich the relation between the nodes of the ontology graph, a functionality for transforming text into concept were needed. The upload functionality is based on a web interface that is able to 
upload a pdf file and associating it with some general information like author, journal, year, volume.
Due to the fact that the more documents are uploaded the better the ontology can be enriched, we code a web spider that is able to use Pubmed site structure to download automatically.
\indent
 Even if the final step of the process, the Enrichment Tool validation, needs a human approval of the concept chosen, this automated download tool helps the boring part of downloading a document from PubMed and uploading to the Chronious system.

The automated download tools is basically a web crawler that periodically use a query string over the Pubmed search functionality, do the HTTP request, parse the html pages and find html tag for having information about the pdf file to download. The code was developed using .NET framework 3.5 being that the CLR provide a library that is able to interact directly with an html page using the DOM (Document Object Model) system. 
\newline
\indent
This allows easily to mimic nearly every interaction that the human user can have with the page like: entering text in the input fields and clicking on a link to simulate the "save as" functionality. 

After the downloading finished successfully the information are stored on a RDBMS database so that only new papers are downloaded and parsed.

One of the problem faced with this approach is the copyright issues that affects the contents of the PubMed database. The greatest part of the articles indexed by PUBMed are protected by copyright, as they come from journals that have copyright agreements, so in most of the case the content of the document cannot be viewed for a user that have no subscription. The problem is not present if we use the upload tool is used inside an institution that have a subscription with PubMed based on the ip address.

However if the Automated Upload Tool is installed on a normal pc downloading document provide an infringement of the PubMed copyright. 
\newline
\indent
To avoid this we decide to use a lower set of documents that are provided by Pubmed for free until the document is published on the journal, for having a first test on the system and to see if the concepts extracted were in line with the ontology we build. To prevent any possible copyright issue, we decide to show to the final user only the DOI of the document so in case the user would like to see the whole document he must open a browser window. This action will shift the copyright from the Chronious system to the final user of the search tools. At the end of the enrichment process the module removes the physical pdf.
\indent
Once the document is uploaded into the repository a Natural Language Processing (NLP) is used to for Information Extraction (IE). 
For implementing the NLP algorithms the GATE \cite{gate} framework will be used. GATE is a leading
infrastructure for developing and deploying software components, that process human language.
Among others it provides a framework, based on JAVA, that implements the architecture and
can be used to embed language processing capabilities in different applications.

GATE supports many document formats like: Plain Text, HTML, SGML, XML, RTF, Email, PDF and
Microsoft Word.

\begin{figure}[htbp]
	\centering
		\includegraphics[width=0.5\textwidth]{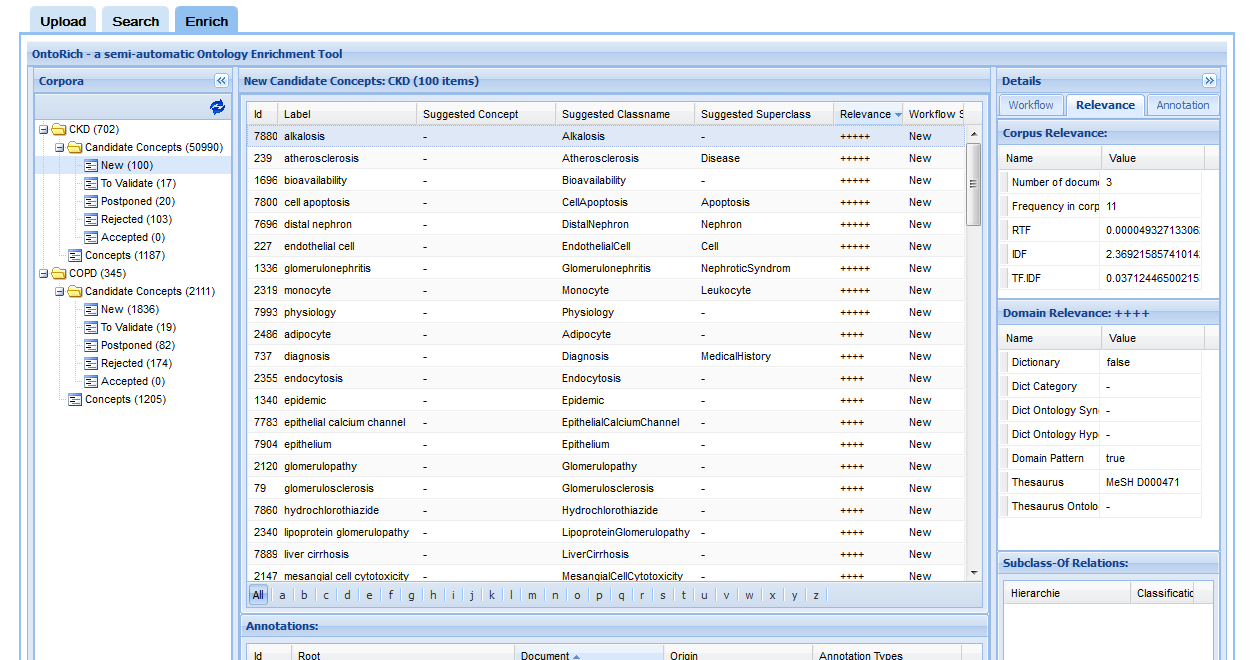}
	\caption{Enrichment Tool}
	\label{enrich}
\end{figure}

Every text is splitted into sentences and words and and every concept extracted is then indexed and associated with a weight that evaluate the correspondence with the other concepts already present into the Chronious ontology. 
\indent
Once this evaluation is done the concept extracted are presented to the human user using a web interface, where the new concepts are shown with their evaluation indexes see Figure \ref{enrich}.

\begin{figure}[htbp]
	\centering
		\includegraphics[width=0.5\textwidth]{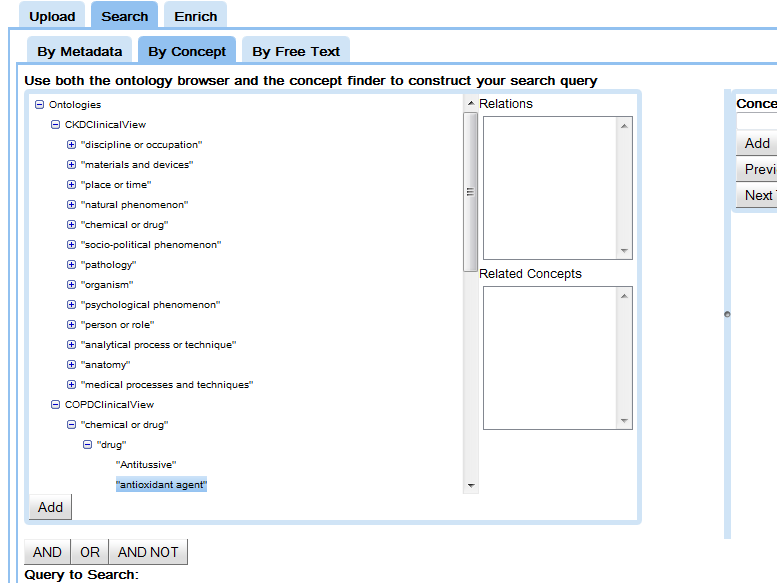}
	\caption{Search Tool}
	\label{search}
\end{figure}

After the whole process finished, a web interface that able to query the ontology for terms evaluation in a structured or free text search form (see Figure \ref{search}) is the last interface. 
So a clinician is able to use it for fast finding literature reference. In addition, the search functionality, uses also the concepts in the ontology so it is possible to have fast reply to query like "abnormal coughing toxicity".

The goodness of the query results should finally be evaluate. Next chapter will detail the criteria used.    
\section{Evaluation}
As we said having another search functionality apart from PubMed raised some murmurs from clinicians, so we need to show that from the perspective of information retrieved, the Chronious Search Tool has some advantages over PubMed.

For doing this, various classifications and criteria for ontology evaluation have been considered in the
literatures \cite{Brank05asurvey}. However, no standard evaluation criteria have been defined so far. Reference literature proposed \cite{Gangemi06modellingontology}
a "three dimensions" classification to evaluate ontology in three categories:
Structural Dimension, Functional Dimension and Usability-Profiling Dimension.
  \indent 
Considering that Structural Dimension and Usability-Profiling Dimension where validated by ontology experts we focused on Functional Dimension with three different tests see Table II:
\begin{table}[htbp]
	\label{tab:table3}
	\centering
	\caption{Features to be tested in Chronious ontology testing}

\begin{tabular}{|l|l|}
\hline 
Test Identifier & Description \\ 
\hline 
TF-B-1 & Agreement Among Experts \\ 
\hline 
TF-B-2 & User-Satisfaction \\ 
\hline 
TF-B-3 & User-Satisfaction and Completeness of Search Result \\ 
\hline 
\end{tabular} 
\end{table}

The questionnaire interview approach can be applied in TF-B-1 to conclude the agreement among
medical experts to evaluate the correctness of the developed Chronious ontologies.
Test participants for this test were medical experts, especially in CKD and/or COPD area.

The black-box test TF-B-2 utilizes the same approach as TF-B-1, the questionnaire interview
approach, to measure the overall satisfaction of the end-user. Because there are three different
search options in the Search Module (Search by Metadata, Search by Concept, and Search by Free
Text), the questionnaire should take all these options into account and compare their search result
quality with each other. The people involved for this interview were the end-user of Chronious
Healthcare Professional GUI, i.e., the medical experts.

In black-box test TF-B-3, Precision $P$ is defined as the number of relevant documents retrieved by a
search $g(r)$ divided by the total number of documents retrieved by that search $N$.
\begin{equation*}
P = \frac{g(r)}{N}
\end{equation*}
while Recall $R$ is defined as the number of relevant documents retrieved by a search $g(r)$ divided by the total number of existing relevant documents $G$ (which should have been retrieved).
\begin{equation*}
R = \frac{g(r)}{G}
\end{equation*}

Precision and Recall are two widely used statistical classifications, especially in information
retrieval domain. Precision can be seen as a measure of exactness or fidelity, whereas Recall is a
measure of completeness.

Usually, Precision and Recall scores are not discussed in isolation.
Instead, either value for one measure are compared for a fixed level at the other measure (e.g.,
Precision at a Recall level of $0.75$), or both are combined into a single measure, such as the $F$-measure \cite{rijsbergen79information}, which is the weighted harmonic mean of Precision and Recall:

\begin{equation*}
F = \frac{(\beta^2+1)\cdot P \cdot R}{(\beta^2\cdot P) + R}
\end{equation*}
Whereby $\beta$ is a value between 0 and 1 reflecting the weighting of Precision vs. Recall.

Before doing an evaluation we uploaded into the repository $1000$ free access documents grabbed by the Automatic Upload Tool from Pubmed using two different query strings 

\begin{itemize}
\renewcommand{\labelitemi}{$\bullet$}
\item{CKD treatment for CKD ontology part.}
\item{COPD treatment limited to year 2008 for COPD ontology part.}
\end{itemize}

Because the document repository of Chronious Search Module and of the PubMed Central
system is artificially identical (constrained with some specific limitations), for same search query,
the two systems should contain the same number of relevant documents to this query, i.e., $G$ for
that query in both systems should be equal. It follows that, no matter which value the
variable $G$ has, the comparison of $F$-measure between these two search systems will only be
influenced by $g(r)$ and $N$ of the search result. Hence, if the value of $g(r)$ and $N$ are available for a
search both with Chronious Conceptual Search option and with PubMed Central system, the $F$-measure
of the search results with both systems can be compared with each other with the help
of their function diagrams (the search query used in both systems must be identical).

Some outputs of our tests can be seen in Table III

\begin{table}[htbp]
	\label{tab:table4}
	\centering
	\caption{Features to be tested in Chronious ontology testing}

\begin{tabular}{|c|c|c|}
\hline 
G & $F$-measure Chronious & $F$-measure Pubmed \\ 
\hline 
500 & 0.724637681 & 0.595238095 \\ 
\hline 
1000 & 0.531914894 & 0.426136364 \\ 
\hline 
2000 & 0.347222222 & 0.27173913 \\ 
\hline 
\end{tabular} 
\end{table}
The table shows that,  the Chronious Conceptual Search option performs better than the PubMed Central system with the search query, no
matter how many relevant documents existed in the repository to this query.
With this approach, although the $F$-measure value of a search result cannot be
estimated for every type of query, the search performance of the Chronious Search Module, however, can be compared
with the search performance of the PubMed Central system, using the $F$-measure.

\section{Conclusion ad further improvements}
Even if the test results seem promising, it is not possible to say that in general the ontology search approach can outperform a search functionality like PubMed one. 
However the good news, is that such kind of approach on storing information, is promising for developing future artificial intelligence systems for application in telemedicine.
Linking concepts like symptoms to drug or caring procedures with relations underline by literature studies can greatly help clinician during their daily routines. 
It would be interesting to evaluate the Conceptual Search on the whole corpus of the Pubmed database even if this would be impossible due to copyright restriction and infrastructure storage possibilities. For sure PubMed will remain the main reference literature search engine, however it is our opinion that having structures information like the one in Chronious ontology, will for sure be an added value.
\newline
\indent
Another interesting task would be the evaluation of a complex free text query search over the two systems. Free text search remains the first approach for searching information. Reflection on how to benefit from ontology structured data in improving outcome for free text search seems is a research problem that require a deeper evaluation.
\newline
\indent
Last we point out that Chronious Enrichment Tool still need to have a human interaction. A still open research task, is the one of having some kinds of automation in inserting the new concepts and relations in the existing ontology, in way to have a sort of unsupervised concepts enrichment that mimics the rule extraction in transaction datasets. We leave these suggestions hoping that the reader will be interested to think about them.  

\bibliographystyle{IEEEtran}
\bibliography{etelemed2012}

\end{document}